\documentstyle[amssymb,12pt,citesort,epsfig]{elsart}

\newcommand{\eq}{\begin{eqnarray}}
\newcommand{\en}{\end{eqnarray}}

\begin{document}

\begin{frontmatter}
\title{Scalar meson and glueball decays within a\\
effective chiral approach} 
\author{F. Giacosa}, 
\author{Th. Gutsche}, 
\author{V.E. Lyubovitskij} 
\author{and Amand Faessler} \\ 
\address{Institut f\"ur Theoretische Physik, Universit\"at T\"ubingen, \\ 
Auf der Morgenstelle 14, D-72076 T\"ubingen, Germany}

\date{\today}
 
\maketitle 
 
\vskip.5cm 

\begin{abstract}
We study the strong and electromagnetic decay properties of scalar mesons 
above 1~GeV within a chiral approach. The scalar-isoscalar states are 
treated as mixed states of quarkonia and glueball configurations.
A fit to the experimental decay rates listed by the Particle Data group 
is performed to extract phenomenological constraints on the nature of the 
scalar resonances. A comparison to other theoretical approaches in the 
scalar meson sector is discussed.


\end{abstract}

\end{frontmatter}

The interpretation of scalar mesons constitutes an unsolved problem of
had\-ro\-nic QCD. Below the mass scale of 2 GeV various scalar states~\cite%
{Eidelman:2004wy} are encountered: the isoscalar resonances $\sigma
=f_{0}(400-1200)$, $f_{0}(980)$, $f_{0}(1370)$, $f_{0}(1500)$ and $%
f_{0}(1710)$, the isovectors $a_{0}(980)$ and $a_{0}(1450) $ and the
isodoublets $K_{0}^{\ast}(800)$ and $K^{\ast}_0(1430).$ The existence of the 
$K_{0}^{\ast}(800)$ is still not well established and omitted from the
summary tables of~\cite{Eidelman:2004wy}. From a theoretical point of view
one expects the scalar quark-antiquark ground state nonet $0^{++}$, a
scalar-isoscalar glueball, which lattice QCD predicts to be the lightest
gluonic meson with a mass between 1.4-1.8 GeV~\cite{Michael:2003ai}, and
possibly other exotic states (non $\bar qq$ states), like e.g. four quark
states or mesonic molecules~\cite{Amsler:2004ps}. Various interpretations of
and assignments for the physical scalar resonances in terms of the expected
theoretical states have been proposed (see, for instance, the review papers~%
\cite{Amsler:2004ps,Close:2002zu} and References therein).

In this work we follow the original assignment of Ref.~\cite{Amsler:1995tu},
where in a minimal scenario the bare quarkonium states $N\equiv \sqrt{1/2}(%
\bar{u}u+\bar{d}d)=\bar{n}n$, $S\equiv \bar{s}s$ and the bare scalar
glueball $G$ mix, resulting in the three scalar-isoscalar resonances $%
f_{0}(1370)$, $f_{0}(1500)$ and $f_{0}(1710)$. Such a mixing scheme has been
previously investigated by many authors, as for example in the lattice study
of~\cite{Lee:1999kv} or within the model approaches of~\cite%
{Close:2001ga,Burakovsky:1998zg,Strohmeier-Presicek:1999yv,Ivanov:1985zw,Giacosa:2004ug}%
. The mesons $a_{0}(1450)$ and $K^{\ast}_0(1430)$ are considered as the $I=1$
and $I=1/2$ quarkonia $J^{PC}=0^{++}$ states. On the other hand, the scalars
below 1 GeV could be interpreted as four quark states or mesonic molecules.

In this Letter, starting from an effective chiral Lagrangian, we perform a
tree-level analysis of the strong and electromagnetic decays of scalar
mesons settled in the energy range between 1 and 2 GeV. Although a chiral
approach cannot be rigorously justified at this energy scale, since loop
corrections can be large, we intend to use this framework as a
phenomenological tool to extract possible glueball-quarkonia mixing
scenarios from the observed decays. This study is in line with the
original application of Ref. \cite{Cirigliano:2003yq}, where detailed
arguments can be found to justify this approach.

{\it Chiral Lagrangian} - The strong and electromagnetic decays of scalar
mesons are based on the effective chiral Lagrangian ${\cal L}_{{\rm eff}}$,
derived in Chiral Perturbation Theory (ChPT)~\cite%
{ChPT,Ecker:1988te,Cirigliano:2003yq}. The Lagrangian involves the nonets of
pseudoscalar (${\cal P}=\sum_{i=0}^{8}P_{i}\lambda _{i}/\sqrt{2}$) and of
scalar mesons (${\cal S}=\sum_{i=0}^{8}S_{i}\lambda _{i}/\sqrt{2}$), the
electromagnetic field and, in addition, a new degree of freedom, the bare
glueball field $G$. We display the lowest order Lagrangian ${\cal L}_{{\rm %
eff}}$, where the starting point is the large $N_{c}$
limit. Corrections to the large $N_{c}$ limit are encoded 
in ${\cal L}_{{\rm mix}}^{P}$ and ${\cal L}_{{\rm mix}}^{S}.$ 
\begin{eqnarray}
\hspace*{-0.85cm}{\cal L}_{{\rm eff}} &=&\frac{F^{2}}{4}\left\langle D_{\mu
}U\,D^{\mu }U^{\dagger }+\chi _{+}\right\rangle +\frac{1}{2}\left\langle
D_{\mu }{\cal S}D^{\mu }{\cal S}-M_{{\cal S}}^{2}{\cal S}^{2}\right\rangle +%
\frac{1}{2}(\partial _{\mu }G\partial ^{\mu }G-M_{G}^{2}G^{2})  \nonumber \\
\hspace*{-0.85cm} &+&c_{d}^{s}\,\left\langle {\cal S}\,u_{\mu }\,u^{\mu
}\,\right\rangle +\,c_{m}^{s}\left\langle {\cal S}\,\chi _{+}\right\rangle
\,\,+\frac{c_{d}^{g}}{\sqrt{3}}G\,\left\langle u_{\mu }\,u^{\mu
}\right\rangle \,+\,\frac{c_{m}^{g}}{\sqrt{3}}\,G\left\langle \chi
_{+}\right\rangle   \nonumber \\
\hspace*{-0.85cm} &+&c_{e}^{s}\,\left\langle {\cal S}\,F_{\mu \nu
}^{+}\,F^{+\,\mu \nu }\right\rangle \,+\,\frac{c_{e}^{g}}{\sqrt{3}}%
\,G\left\langle \,F_{\mu \nu }^{+}\,F^{+\,\mu \nu }\right\rangle \,+\,{\cal L%
}_{{\rm mix}}^{P}+\,{\cal L}_{{\rm mix}}^{S}\,.  \label{L_eff}
\end{eqnarray}%
Here the symbol $\left\langle ...\right\rangle $ denotes the trace over
flavor matrices. The constants $c_{d}^{s}$, $c_{m}^{s}$, $c_{d}^{g}$, $%
c_{m}^{g}$, $c_{e}^{s}$ and $c_{e}^{g}$ define the coupling of scalar fields
and the bare glueball to pseudoscalar mesons and photons, respectively (the
coupling constant $c_{e}^{g}$ is supposed to be suppressed because gluons do
not couple directly to photons. However, an intermediate state 
of two vector mesons can in the framework of vector meson dominance
generate a coupling of the glueball to the two-photon final state;
this coupling is however not considered in the numerical analysis). 
The field $G$ is introduced as
an extra flavor-singlet field in the effective lagrangian. 
We use the
standard notation for the basic blocks of the ChPT Lagrangian~\cite{ChPT}: $%
U=u^{2}=\exp (i{\cal P}\sqrt{2}/F)$ is the chiral field collecting
pseudoscalar fields in the exponential parametrization, $D_{\mu }$ denotes
the chiral and gauge-invariant derivative, $u_{\mu }=iu^{\dagger }D_{\mu
}Uu^{\dagger }$, \hspace*{0.2cm} $\chi _{\pm }=u^{\dagger }\chi u^{\dagger
}\pm u\chi ^{\dagger }u,\hspace*{0.2cm}\chi =2B(s+ip),\,\,\,s={\cal M}%
+\ldots \,$ and $F_{\mu \nu }^{+}\,=\,(\,u^{\dagger }F_{\mu \nu }Qu+uF_{\mu
\nu }Qu^{\dagger }\,)\,,$ where $F_{\mu \nu }$ is the stress tensor of the
electromagnetic field; $Q={\rm diag}\{2/3,-1/3,-1/3\}$ and ${\cal M}={\rm %
diag}\{\hat{m},\hat{m},m_{s}\}$ are the charge and the mass matrix of
current quarks, respectively (we restrict to the isospin symmetry limit with 
$m_{u}=m_{d}=\hat{m}$); $B$ is the quark vacuum condensate parameter. Then
the masses of the pseudoscalar mesons in the leading order of the chiral
expansion are given by $M_{\pi }^{2}=2\hat{m}B\,\,,M_{K}^{2}=(\hat{m}%
+m_{s})B\,\,,\,M_{\eta ^{8}}^{2}=(2/3)(\hat{m}+2m_{s})B\,.$

We intend to employ the Lagrangian (\ref{L_eff}) for the tree-level
calculation of the two-pseudoscalar decays of scalar mesons. At the
energy scale of interest, $E \sim M_{{\cal S}}~\sim~1.5$ GeV, a
calculation of loops and an application of the power counting rules
are not rigorously justified. The aim of the present approach is 
a phenomenological study of scalar meson physics, for which a tree-level 
calculation represents a useful analysis of the issue.

The term ${\cal L}_{{\rm mix}}^{P}\sim \eta ^{0}\eta ^{8}$ gives rise to the
singlet-octet mixing in the pseudoscalar and the scalar multiplets. The
physical states $\eta $ and $\eta ^{\prime }$ are given by 
\[
\eta ^{0}\,=\,\eta ^{\prime }\,\cos \theta _{P}\,-\,\eta \,\sin \theta
_{P}\,,\hspace*{0.25cm}\eta ^{8}\,=\,\eta ^{\prime }\,\sin \theta _{P}\,\
+\,\eta \,\cos \theta _{P}\,, 
\]%
where $\theta _{P}$ is the pseudoscalar mixing angle. We follow the standard
procedure~\cite{Ecker:1988te,Cirigliano:2003yq,Venugopal:1998fq} and
diagonalize the corresponding $\eta ^{0}$-$\eta ^{8}$ mass matrix to obtain
the masses of $\eta $ and $\eta ^{\prime }$. By using $M_{\pi }=139.57$ MeV, 
$M_{K}=493.677$ MeV (the physical charged pion and kaon masses), $M_{\eta
}=547.75$ MeV and $M_{\eta ^{\prime }}=957.78$ MeV the mixing angle is
determined as $\theta _{P}=-9.95^{\circ }$, which corresponds to the
tree-level result (see details in Ref.~\cite{Venugopal:1998fq}). Higher
order corrections in ChPT cause a doubling of the absolute value of the
pseudoscalar mixing angle~\cite{Venugopal:1998fq}); in our work we restrict
to the tree-level evaluation, we therefore consistently use the
corresponding tree-level result of $\theta _{P}=-9.95^{\circ }\,\ $(in the
present approach we do not include the neutral pion when considering
mixing in the pseudoscalar sector, because we work in the isospin limit;
this mixing is small, and can be safely neglected when studying the decay of
scalar resonances into two pseudoscalars). Similarly, for all pseudoscalar
mesons we use the unified leptonic decay constant $F$, which is identified
with the pion decay constant $F=F_{\pi }=92.4$~MeV. A more accurate analysis
including higher orders should use the individual couplings of the
pseudoscalar mesons (for a detailed discussion see Refs.~\cite{Gasser:1984gg}%
).

{\it Glueball-Quarkonia Mixing} - We briefly discuss the mixing of the
glueball with the scalar-isoscalar quarkonia states. The singlet $(S^{0})$
and octet $(S^{8})$ scalar quarkonia states are defined in terms of the
nonstrange $N$ and strange $S$ components with the respective flavor content 
$[\bar{u}u+\bar{d}d]/\sqrt{2}$ and $\bar{s}s$ as 
\begin{equation}
S^{0}\,=\,\sqrt{2/3}\,N\,+\,\sqrt{1/3}\,S\,,\hspace*{0.25cm}S^{8}\,=\,\sqrt{%
1/3}\,N-\,\sqrt{2/3}\,S\,.
\end{equation}%
In (\ref{L_eff}) the scalar nonet states have the same mass $M_{{\cal S}},$
corresponding to the flavor and large $N_{c}$ limits. Deviations from this
configuration (i.e. from the large $N_{c}$ limit) are encoded in $\,{\cal L}_{%
{\rm mix}}^{S}$ including quarkonia-glueball mixing leading to different
masses for the scalar mesons (in~\cite{Cirigliano:2003yq} this is explicitly
shown for the scalar nonet, but no glueball is included; the glueball
mixing is also a deviation from large $N_{c},$ since in this limit the
glueball and the quarkonia states decouple). 
As a result, the scalar-isoscalar
sector can be described by the most general Klein-Gordon (KG) Lagrangian
including mixing among the $N$, $S$ and the bare glueball $G$ configurations~%
\cite%
{Amsler:1995tu,Lee:1999kv,Close:2001ga,Burakovsky:1998zg,Strohmeier-Presicek:1999yv}%
: 
\begin{equation}
{\cal L}_{KG}=-\frac{1}{2}\sum\limits_{\Phi =N,G,S}\,\Phi \lbrack \Box
+M_{\Phi }^{2}]\Phi -fGS-\sqrt{2}frGN-\varepsilon NS\,,  \label{L_mix}
\end{equation}
where $\Box =\partial ^{\mu }\partial _{\mu }$. The parameter $f$ is the
quarkonia-glueball mixing strength, analogous to the parameter $z$ of Refs.
\cite{Amsler:1995tu,Lee:1999kv,Close:2001ga,Burakovsky:1998zg,Strohmeier-Presicek:1999yv};
$z$ refers to the quantum mechanical case where the mass matrix is linear
in the bare mass terms, $f$ in turn is related to the the quadratic
Klein-Gordon case. The connection between $f$ and $z$ discussed in Refs.~%
\cite{Burakovsky:1998zg,Giacosa:2004ug} leads to the approximate relation $%
f\simeq 2zM_{G}\,.$ If $r=1$, then the glueball is flavor blind and mixes
only with the quarkonium flavor singlet $S^{0}$ (this is the case in Refs.~%
\cite{Amsler:1995tu,Burakovsky:1998zg,Strohmeier-Presicek:1999yv}); a value $%
r\neq 1$ takes into account a possible deviation from this limit. However,
as deduced from the analyses of Refs.~\cite%
{Lee:1999kv,Close:2001ga,Giacosa:2004ug} $r$ is believed to be close to
unity. In the following we will use the limit $r=1,$ i.e. we restrict to a
flavor blind mixing.

The parameter $\varepsilon $ induces a direct mixing between $N$ and $S$
quarkonia states. This effect is neglected in~\cite%
{Amsler:1995tu,Burakovsky:1998zg,Close:2001ga,Strohmeier-Presicek:1999yv},
where flavor mixing is considered a higher-order effect. However, a
substantial $N$-$S$ mixing in the scalar sector is the starting point of the
analysis of Refs.~\cite{Minkowski:2002nf,Minkowski:1998mf}. The origin of
quarkonia flavor mixing is, according to~\cite%
{Minkowski:2002nf,Minkowski:1998mf}, connected to instantons as in the
pseudoscalar channel, but with opposite sign (see also~\cite{Klempt:1995ku}%
): the mixed physical fields are predicted to be a higher lying state of
flavor structure $[N\sqrt{2}-S]/\sqrt{3}$ and a lower one with $[N+S\sqrt{2}%
]/\sqrt{3}$. Here we study the case $\varepsilon \neq 0,$ more precisely $%
\varepsilon >0,$ which leads to the same phase structure as in Ref.~\cite%
{Minkowski:2002nf,Minkowski:1998mf}, but the quantitative results and
interpretation will differ. The physical scalar states $\left\vert
i\right\rangle $ are identified with $i=f_{1}\equiv f_{0}(1370)$, $%
f_{2}\equiv $ $f_{0}(1500)$ and $f_{3}\equiv f_{0}(1710)$, which are set up
as linear combinations of the bare states: $\left\vert i\right\rangle
=B^{iN}|N\rangle +B^{iG}\left\vert G\right\rangle +B^{iS}\left\vert
S\right\rangle $. The amplitudes $B^{ij}$ are the elements of a matrix $B$
which diagonalizes the mass matrix of bare states including mixing, which in
turn gives rise to the mass matrix of physical states $\Omega ^{\prime }=%
{\rm diag}\{M_{f_{1}}^{2},M_{f_{2}}^{2},M_{f_{3}}^{2}\}\,.$

{\it Results for suppressed glueball decays} - In the following we determine
a best fit of the parameters entering in Eqs.~(\ref{L_eff}) and (\ref{L_mix}%
) to the experimental averages of masses and decay modes listed in Ref.~\cite%
{Eidelman:2004wy}. Decay rates resulting from Eq.~(\ref{L_eff}) are
evaluated on the tree-level. First we analyze the case of a non-decaying
glueball, i.e. $c_{d}^{g}\,=c_{m}^{g}=c_{e}^{g}=0$. In this scheme the
decays are dominated by the quarkonia components~\cite{Amsler:1995tu} in
line with large $N_{c}$ arguments. The phenomenological analysis of Ref.~%
\cite{Strohmeier-Presicek:1999yv} confirmed this trend, but the recent fit
of~\cite{Close:2001ga} shows a strong contribution by the direct decays of
the glueball configuration. The strength of the glueball decays remains a
point to clarify. Our results for suppressed glueball decays are given in
Table 1 in comparison to the data of Ref.~\cite{Eidelman:2004wy}. The only
accepted average not included in the fit is $\Gamma _{f_{2}\rightarrow \eta
\eta ^{\prime }}$ for the reason that the decay channel $\eta \eta ^{\prime }
$ is produced at threshold, therefore a significant distortion due to the
finite width of the state is expected. The state $f_{0}(1710)$ has only been
observed in the decays into two pseudoscalar mesons \cite{Eidelman:2004wy}.
The decay into the final state $4\pi $, which can be fed by higher meson
resonances, is suppressed~\cite{Barberis:2000cd}. We therefore impose the
additional condition that the sum of partial decay widths into two
pseudoscalar mesons $\left( \Gamma _{f_{3}}\right) _{2P}$ saturates the
total width $\left( \Gamma _{f_{3}}\right) _{tot}$ with $\left( \Gamma
_{f_{3}}\right) _{2P}=\left( \Gamma _{f_{3}}\right) _{tot}$ as indicated in
Table 1. Such a constraint is necessary to obtain meaningful total decay
widths: without this condition on the full width a minimum for $\chi ^{2}$
is obtained where $(\Gamma _{f_{3}})_{tot}$ is larger than $1$ GeV, a
clearly unacceptable solution. A local minimum for $\chi _{tot}^{2}$ is
found for the set of parameters 
\begin{eqnarray}
\hspace*{-0.75cm}M_{N} &=&1.455~{\rm GeV},~M_{G}=1.490~{\rm GeV}%
,~M_{S}=1.697~{\rm GeV},~f=0.065~{\rm GeV}^{2},  \nonumber \\
\hspace*{-0.75cm}\varepsilon  &=&0.211~{\rm GeV}^{2},~c_{d}^{s}=8.48~{\rm MeV%
},~c_{m}^{s}=2.59~{\rm MeV}~\chi _{tot}^{2}=29.01\,.  \label{fitpar}
\end{eqnarray}

\begin{center}
{\bf Table 1.} Mass and decay properties of scalar $f_0$ mesons.

\begin{tabular}{llll}
Quantity & Exp & Theory & $\chi _{i}^{2}$ \\ 
$M_{f_{1}}$ \thinspace ({\rm MeV}) & $1350$ $\pm 150$ & $1417$ & $0.202$ \\ 
$M_{f_{2}}$ \thinspace ({\rm MeV}) & $1507\pm 5$ & $1507$ & $\sim 0$ \\ 
$M_{f_{3}}$ \thinspace ({\rm MeV}) & $1714\pm 5$ & $1714$ & $0.003$ \\ 
$\Gamma _{f_{2}\rightarrow \pi \pi }$ \thinspace \thinspace ({\rm MeV}) & $%
38.0\pm 4.6$ & $38.52$ & $0.011$ \\ 
$\Gamma _{f_{2}\rightarrow \bar{K}K}$ \thinspace \thinspace ({\rm MeV}) & $%
9.4\pm 1.7$ & $10.36$ & $0.322$ \\ 
$\Gamma _{f_{2}\rightarrow \eta \eta }$ \thinspace ({\rm MeV}) & $5.6\pm 1.4$
& $1.90$ & $8.109$ \\ 
$\Gamma _{f_{3}\rightarrow \pi \pi }/\Gamma _{f_{3}\rightarrow \bar{K}K}$ & $%
0.20\pm 0.06$ & $0.212$ & $0.036$ \\ 
$\Gamma _{f_{3}\rightarrow \eta \eta }/\Gamma _{f_{3}\rightarrow \bar{K}K}$
& $0.48\pm 0.15$ & $0.249$ & $2.446$ \\ 
$\Gamma _{a_{0}\rightarrow \bar{K}K}/\Gamma _{a_{0}\rightarrow \pi \eta }$ & 
$0.88\pm 0.23$ & $0.838$ & $0.032$ \\ 
$\Gamma _{a_{0}\rightarrow \pi \eta ^{\prime }}/\Gamma _{a_{0}\rightarrow
\pi \eta }$ & $0.35\pm 0.16$ & $0.288$ & $0.150$ \\ 
$\Gamma _{K^{\ast }\rightarrow K\pi }$ \thinspace ({\rm MeV}) & $273\pm 51$
& $59.10$ & $17.590$ \\ 
$\left( \Gamma _{f_{3}}\right) _{2P}$ \thinspace ({\rm MeV}) & $140\pm 10$ & 
$143.27$ & $0.110$ \\ 
$\chi _{tot}^{2}$ & - & - & $29.01$%
\end{tabular}
\end{center}

{\it Implications of the fit} - The bare non-strange quarkonia field $N$ has
a mass of $M_{N}=1.455$ {\rm GeV} which is, as desired, similar to the scale
set by the isotriplet combination $a_{0}(1450)$ with a mass of $%
M_{a_{0}}=1.474\pm 0.019$ {\rm GeV} \cite{Eidelman:2004wy}. The mass of the
bare glueball $M_{G}=1.490$ {\rm GeV} is in agreement with the lattice
results~\cite{Michael:2003ai} and with the phenomenological analyses of~\cite%
{Amsler:1995tu,Close:2001ga,Strohmeier-Presicek:1999yv}. The bare state $S$
has a mass of $M_{S}=1.697$ {\rm GeV}, which is about $\sim 200$ MeV heavier
than the $N$ state, an acceptable mass difference like in the tensor meson
nonet.

For the glueball-quarkonia mixing parameter we get $f=0.065$ GeV$^{2}$,
which by the approximate relation $f\simeq 2zM_{G}$~\cite%
{Burakovsky:1998zg,Giacosa:2004ug} corresponds to $z\simeq 21.8$ MeV. The
results of Refs.~\cite%
{Close:2001ga,Strohmeier-Presicek:1999yv,Giacosa:2004ug} are $z=85\pm 10$
MeV, $z=80$ MeV and $z\simeq 62$ MeV, respectively, i.e. of the same order,
but larger. The introduction of additional flavor mixing between the
quarkonia configurations in the fit, as done here, leads to a reduction of
the strength parameter $f$. The lattice result of Ref.~\cite{Lee:1999kv}
with $43\pm 31$ MeV is in agreement with the present evaluation, but has a
large uncertainty.

The flavor mixing parameter resulting from the fit is $\varepsilon =0.211$
GeV$^{2}$. In the limit $f=0$ the mixed physical states are $\left\vert
f_{1}\right\rangle =0.97\left\vert N\right\rangle +0.26\left\vert
S\right\rangle $ and $\left\vert f_{3}\right\rangle =-0.26\left\vert
N\right\rangle +0.97\left\vert S\right\rangle $ (and, of course, $\left\vert
f_{2}\right\rangle =\left\vert G\right\rangle $). The phase structure of the
mixed states is, as discussed previously, as in \cite%
{Minkowski:2002nf,Minkowski:1998mf,Klempt:1995ku}. But here the strength of
flavor mixing is smaller, resulting in mixed states which are dominantly $N$
or $S$. The influence however of (an even small) flavor mixing in strong and
electromagnetic decays may be non-negligible.

The mixing matrix $B$ relating the physical to the bare states in the
present fit is expressed as: 
\begin{equation}
\,\left( 
\begin{array}{l}
\left\vert f_{1}\right\rangle \equiv \left\vert f_{0}(1370)\right\rangle  \\ 
\left\vert f_{2}\right\rangle \equiv \left\vert f_{0}(1500)\right\rangle  \\ 
\left\vert f_{3}\right\rangle \equiv \left\vert f_{0}(1710)\right\rangle 
\end{array}%
\right) =\left( 
\begin{array}{lll}
0.86 & 0.45 & 0.24 \\ 
-0.45 & 0.89 & -0.06 \\ 
-0.24 & -0.06 & 0.97%
\end{array}%
\right) \left( 
\begin{array}{l}
\left\vert N\right\rangle \equiv \left\vert \bar{n}n\right\rangle  \\ 
\left\vert G\right\rangle \equiv \left\vert gg\right\rangle  \\ 
\left\vert S\right\rangle \equiv \left\vert \bar{s}s\right\rangle 
\end{array}%
\right) .  \label{mixingmat}
\end{equation}%
The physical resonances are dominated by the diagonal bare components,
qualitatively in line with Refs.~\cite%
{Amsler:2004ps,Amsler:1995tu,Close:2001ga,Strohmeier-Presicek:1999yv}. Since
the glueball does not contribute to the decay, the relative phase with
respect to the quarkonia components is at this stage irrelevant; by
inverting $f\rightarrow -f$ we would find the same results for the decays,
but opposite glueball-quarkonia phases. In turn, the relative phases of the $%
N$ and $S$ components are not symmetric under $\varepsilon \rightarrow
-\varepsilon $; as discussed above, in $f_{0}(1370)$ they are in phase,
while in $f_{0}(1710)$ they are out of phase. The state $\left\vert
f_{0}(1500)\right\rangle $ behaves like a $N$ state with a decreased width,
while the S component is small \cite%
{Strohmeier-Presicek:1999yv,Amsler:2002ey}. Thus the decay into $K\bar{K}$
is smaller than the $\pi \pi $ channel. The $N$ and $S$ state components are
in phase contrary to the results of \cite%
{Amsler:1995tu,Strohmeier-Presicek:1999yv}.

The experimental uncertainties of the $f_{0}(1370)$ resonance are large, no
average or fit is presented in \cite{Eidelman:2004wy}. The main problem
connected with this resonance is its large width $(200$-$500$ $MeV)$ and its
partial overlap with the broad low-lying $\sigma \equiv f_{0}(400-1200).$
However, the results from WA102~\cite{Barberis:2000cd} indicate a large $N$
component in its wave function; the results from Crystal Barrel (summarized
in~\cite{Amsler:1997up} and subsequently analyzed in~\cite{Abele:2001pv})
confirm such a trend (see also~\cite{Thoma:2003in} for a recent review).

\begin{center}
{\bf Table 2.} Decays of $f_{1}=f_{0}(1370)$.

\begin{tabular}{lll}
Quantity & Exp (WA102) & Theory \\ 
$\Gamma _{f_{1}\rightarrow \bar{K}K}/\Gamma _{f_{1}\rightarrow \pi \pi }$ & $%
0.46\pm 0.19$ & $0.34$ \\ 
$\Gamma _{f_{1}\rightarrow \eta \eta }/\Gamma _{f_{1}\rightarrow \pi \pi }$
& $0.16\pm 0.07$ & $0.06$ \\ 
$\left( \Gamma _{f_{1}}\right) _{2P}$ ({\rm MeV}) & \textquotedblright
small\textquotedblright  & $166$%
\end{tabular}
\end{center}

Predictions for the two-pseudoscalar decay modes are in acceptable agreement
with the results of WA102 as shown in Table 2. The measured ratio $\Gamma
_{f_{1}\rightarrow 4\pi }/\Gamma_{f_{1} \rightarrow \pi \pi
}=34.0_{-9}^{+22} $ \cite{Barberis:2000cd}, although the errors are large,
points to a dominant $4\pi $ contribution to the total width. Our prediction
gives however a sizable contribution of the two-pseudoscalar decay mode.

An analysis by Crystal Barrel~\cite{Abele:2001pv} also indicates sizable
partial decay widths of the $4\pi $ decay channels: $\Gamma
_{f_{1}\rightarrow \sigma \sigma }=120.5\pm 65$ MeV and $\Gamma
_{f_{1}\rightarrow \rho \rho }=62.2\pm 28.8$ MeV. The same analysis gives
the following two-pseudoscalar partial widths~\cite%
{Abele:2001pv,Thoma:2003in}: $\Gamma _{f_{1}\rightarrow \pi \pi }=21.7\pm 9.9
$ {\rm MeV}\thinspace, $\Gamma _{f_{1}\rightarrow K\bar{K}}=(7.9\pm 2.7$ 
{\rm MeV}) to ($21.2\pm 7.2$ MeV) \thinspace , $\Gamma _{f_{1}\rightarrow
\eta \eta }=0.41\pm 0.27$ MeV. On the contrary, the analysis of \cite%
{Bugg:1996ki} reports the ratio $\Gamma _{f_{1}\rightarrow \pi \pi }/(\Gamma
_{f_{1}})_{tot}=0.26\pm 0.09,$ which is in disagreement with the previously
discussed experimental results.

In the original work~\cite{Amsler:1995tu}, a quarkonium state $N$ has a very
large two-pseu\-do\-sca\-lar width with $(\Gamma _{N})_{2P}\sim 500$ MeV.
Since the $f_{0}(1370)$ is interpreted as a dominant $N$ state \cite%
{Amsler:1995tu} one also expects a large value for $(\Gamma _{f_{1}})_{2P}$.
A sizable value for $(\Gamma _{N})_{2P}$ is also predicted in \cite%
{Strohmeier-Presicek:1999yv}; for the mixed state $f_{0}(1370)$ one has $%
(\Gamma _{f_{1}})_{2P}=115.7$ {\rm MeV} comparable to the present result. In
a recent fit \cite{Close:2001ga} small two-pseudoscalar partial decay widths
are obtained by the following mechanism: for $f_{0}(1370)$ the glueball
component gives large decay amplitudes which interfere destructively with
the contribution of the $N$ component. As a result \cite{Close:2001ga}, the
two-pseudoscalar decay width $(\Gamma _{f_{1}})_{2P}$ is smaller than $%
(\Gamma _{f_{2}})_{2P}$ with $(\Gamma _{f_{2}})_{2P}/(\Gamma
_{f_{1}})_{2P}=10.0\pm 3.0;$ at the same time $(\Gamma
_{f_{3}})_{2P}/(\Gamma _{f_{2}})_{2P}=0.7\pm 0.2.$ In the present fit with a
suppressed glueball decay we have the following decay widths into two
pseudoscalar pairs: $(\Gamma _{f_{1}})_{2P}=165.79$ MeV $>(\Gamma
_{f_{3}})_{2P}=143.27$ MeV $>(\Gamma _{f_{2}})_{2P}=50.82$ MeV. The
experimental situation concerning the absolute scale of the partial decay
widths into pseudoscalars remains unclear, but the explicit values
constitute a key test of mixing scenarios.

The theoretical partial widths of the $f_{0}(1500)$ are in good agreement
with the data (see Table 1) apart from a slight underestimate of the $2\eta $
channel. We also obtain $\Gamma _{f_{2}\rightarrow \eta \eta ^{\prime
}}=0.036$ MeV as compared to the experimental value of $\Gamma
_{f_{2}\rightarrow \eta \eta ^{\prime }}=2\pm 1$ MeV. Taking into account
the finite width of the resonance will lead to an increase of the
theoretical value.

For the decays of $f_{0}(1710)$ we summarize our results compared to the
data of WA102~\cite{Barberis:2000cd} in Table 3.

\begin{center}
{\bf Table 3.} Decays of $f_{3}=f_{0}(1710)$.

\begin{tabular}{lll}
Quantity & Exp (WA102) & Theory \\ 
$\Gamma_{f_{3}\rightarrow \bar{K}K}/\Gamma_{f_{3}\rightarrow \pi \pi } $ & $%
5.0\pm 0.7$ & $4.70$ \\ 
$\Gamma_{f_{3}\rightarrow \eta \eta }/\Gamma_{f_{3}\rightarrow \pi \pi }$ & $%
2.4\pm 0.6$ & $1.17$ \\ 
$\Gamma_{f_{3}\rightarrow \eta \eta^{\prime }}/\Gamma_{f_{3}\rightarrow \pi
\pi }$ & $<0.18$ & $1.59$ \\ 
$\left( \Gamma_{f_{3}}\right)_{2P}$ ({\rm MeV}) & ''dominant'' & $143.27$%
\end{tabular}
\end{center}

The first two ratios, already included in the fit of Table 2, can be
reproduced. The theoretical ratio $\Gamma_{f_{3}\rightarrow \eta\eta^{\prime
}}/\Gamma_{f_{3}\rightarrow \pi \pi}$, which is not included in \cite%
{Eidelman:2004wy}, is in complete disagreement with the WA102 result. The
dominance of the $\eta\eta^{\prime}$ mode over $\pi\pi$ is a solid
prediction, which does not depend very much on the choice of parameters. A
confirmation of the experimental result could possibly hint at a sizable
role of direct glueball decay.

The ratios of two-pseudoscalar decay modes of $a_{0}(1450)$, included in the
fit of Table 2, are well reproduced. The prediction for the two-pseudoscalar
width of $(\Gamma _{a_{0}})_{2P}=\ 84.26$ MeV is smaller than the total
width of $265\pm 13$ MeV. However, the experimental ratio ($\Gamma
_{a_{0}\rightarrow \omega \pi \pi }/\Gamma _{a_{0}\rightarrow \pi \eta })$
is not known: no average or fit is listed in~\cite{Eidelman:2004wy}. The
experimental value from~\cite{Baker:2003jh}, which is $10.7\pm 2.3$, would
imply a dominant $\omega \pi \pi $ mode and in turn rather small
two-pseudoscalar partial decay widths. This finding is in disagreement with
the results of~\cite{Amsler:1995tu,Gobbi:1993au}. In~\cite{Amsler:1995tu} a
value of $(\Gamma _{a_{0}})_{2P}=\ 390\pm 110$ MeV is found, in the work of~%
\cite{Gobbi:1993au} one has $\left( \Gamma _{a_{0}}\right) _{2P}=\ 420$-$940$
MeV for $a_{0}$ masses in the range of $1200$-$1400$ MeV. For our value for $%
(\Gamma _{a_{0}})_{2P}$ we obtain the estimate: $\Gamma _{a_{0}\rightarrow
\omega \pi \pi }/\Gamma _{a_{0}\rightarrow \pi \eta }=[(\Gamma
_{a_{0}})_{tot}-(\Gamma _{a_{0}})_{2P}]/\ \Gamma _{a_{0}\rightarrow \pi \eta
}\sim 4.5\,.$

Our result for $\Gamma _{K^{\ast }\rightarrow K\pi }$ underestimates the
experimental value by about a factor of 5 (see Table 1). Furthermore, for
the additional $K\eta $ decay channel we get $\Gamma _{K^{\ast }\rightarrow
K\eta }/\Gamma _{K^{\ast }\rightarrow \pi K}=0.026$.

In~\cite{Gobbi:1993au} a value of $\Gamma _{K^{\ast }\rightarrow K\pi }=340$
MeV is predicted, but, as discussed above, $\left( \Gamma _{a_{0}}\right)
_{2P}$ is of the order of $1$ GeV, much larger than the full width.
Similarly, in~\cite{Amsler:1995tu} with $\Gamma _{K^{\ast }\rightarrow K\pi
}=200\pm 20$ and $(\Gamma _{a_{0}})_{2P}=390\pm 110$ MeV the first result
underestimates while the second overshoots the experimental value. A full
analysis in the $^{3}P_{0}$ model~\cite{Ackleh:1996yt} results in $\Gamma
_{K^{\ast }\rightarrow \pi K}=166$~MeV, $\Gamma _{N\rightarrow \pi \pi }=271$%
~MeV. Unfortunately, 
the resonance $a_{0}(1450)$ is not discussed in~\cite{Ackleh:1996yt}. The
authors of \cite{Ackleh:1996yt} also tried to fix $\Gamma _{K^{\ast
}\rightarrow \pi K}$ to its experimental value, and then calculate the $2\pi 
$ partial width of a $N$ state, obtaining $\Gamma _{N\rightarrow \pi \pi
}\sim 450$ MeV. The last result implies a very large two-pseudoscalar and
full width for a $N$ state. 
A full experimental determination of all relevant decay modes
involving $a_{0}(1450)$ and $K^{\ast}_0(1430)$ would certainly help to
clarify this issue.

If $f_{0}(1370)$ is dominantly $\bar{n}n,$ as in~\cite%
{Amsler:1995tu,Close:2001ga,Lee:1999kv,Burakovsky:1998zg,Strohmeier-Presicek:1999yv,Giacosa:2004ug}%
, there is, as discussed above, an incompatibility of the present
experimental small two-pseudoscalar partial decay widths~\cite{Abele:2001pv}
and various model calculations. At the same time, the consistent
understanding of the isodoublet states $K^{\ast}(1450)$, the isovectors $%
a_{0}(1450)$ is still incomplete.

Proceeding as in a~\cite{Ackleh:1996yt}, that is fitting $%
\Gamma_{K^{\ast}\rightarrow \pi K}$ (together with the ratios of $%
a_{0}(1450) $) we find $c_{d}^{s}=0.0180$ GeV (larger than a factor 2 when
compared to the previous fit in Table 1) and $c_{m}^{s}=0.0073$ GeV. With
these values one has $\Gamma _{K^{\ast}\rightarrow \pi K}=281$ MeV, $%
\Gamma_{a_{0}\rightarrow K\bar{K}}/\Gamma_{a_{0}\rightarrow \pi \eta} = 0.88$
and $\Gamma_{a_{0}\rightarrow \pi \eta^{\prime }}/\Gamma_{a_{0} \rightarrow
\pi\eta }=0.30$ and $\Gamma_{a_{0}\rightarrow 2P}=381.17$ MeV. The
corresponding decay width of the $N$ state into two pseudoscalars is $\Gamma
_{N\rightarrow 2P}=937$ MeV (for a mass of $M_{N}=1.448$ GeV). We then find
similar results as in~\cite{Amsler:1995tu,Ackleh:1996yt}, implying that the
discussed trend is rather model independent. Fitting the remaining free
parameters to the experimental data, a $\chi^2$ minimum results in: $%
M_{N}=1.453$ GeV \thinspace, $M_{G}=1.504$ GeV \thinspace, $M_{S}=1.698$
GeV \thinspace, $f=0.0281$ GeV $^{2}$\thinspace, $\varepsilon =0.209$ GeV$%
^{2}\,.$ The mixing parameter $f$ is much smaller than in the full fit, in
disagreement with the results of~\cite%
{Amsler:1995tu,Close:2001ga,Strohmeier-Presicek:1999yv}. As a further
consequence the state $\left| f_{1}\right\rangle \simeq \left|
N\right\rangle $ has a very large two-pseudoscalar width of about $1$ GeV.
At the same time $\left|f_{3}\right\rangle \simeq \left| S\right\rangle ,$
which also has a large (and not acceptable) two-pseudoscalar width of $\sim
700$~MeV.

As a further consequence we discuss the two-photon decay rates of the scalar
resonances. We assume that the coupling $c_{e}^{g}$ is suppressed with
respect to $c_{e}^{s},$ i.e. we set the glueball-photon coupling $c_{e}^{g}$
to zero. The ratios of radiative decay widths as a prediction of the fit
are: 
\begin{equation}
\Gamma _{f_{1}\rightarrow 2\gamma }:\Gamma _{f_{2}\rightarrow 2\gamma
}:\Gamma _{f_{3}\rightarrow 2\gamma }:\Gamma _{a_{0}^{0}\rightarrow 2\gamma
}=1:0.367:0.004:0.491\,,
\end{equation}%
which are independent of the coupling $c_{e}^{s}$. The result for $\Gamma
_{f_{2}\rightarrow 2\gamma }/\Gamma _{f_{1}\rightarrow 2\gamma }$ is in
qualitative agreement with the results of~\cite{Close:2001ga,Giacosa:2004ug}%
. The ratio $\Gamma _{f_{3}\rightarrow 2\gamma }/\Gamma _{f_{1}\rightarrow
2\gamma }$, however, is considerably smaller than in the previous works. The
suppression of $\Gamma _{f_{3}\rightarrow 2\gamma }$ originates from the
destructive interference between the $N$ and $S$ components, which in turn
is traced to the flavor mixing with $\varepsilon >0$ in accord with the
phases of~\cite{Minkowski:2002nf,Minkowski:1998mf}. Another interesting
prediction is the ratio $\Gamma _{a_{0}^{0}\rightarrow 2\gamma }/\Gamma
_{f_{1}\rightarrow 2\gamma }$, which is relatively large.

The experimental status of the two-photon decays is still incomplete. For
the $f_{0}(1370)$ two values are indicated in PDG2000~\cite{Groom:in}: $%
3.8\pm 1.5$ ${\rm keV}$ and $5.4\pm 2.3$ keV. However, it is not clear if
the two-photon signal comes from the $f_{0}(1370)$ or from the high mass end
of the broad $f_{0}(400-1200).$ The PDG currently~\cite{Eidelman:2004wy}
seems to favor this last possibility, but the data could also be valid for
the $f_{0}(1370).$ We therefore interpret the two experimental values as an
upper limit for the two-photon decay width of the $f_{0}(1370).$ Signals for
two-photon decays of $f_{0}(1500)$ and $f_{0}(1710)$ have not yet been seen;
the following upper limits are reported~\cite{Eidelman:2004wy}: 
\begin{eqnarray}
\Gamma _{f_{0}(1500)\rightarrow 2\gamma }(\Gamma _{f_{0}(1500)\rightarrow
\pi \pi }/\Gamma _{f_{0}(1500)tot}) &<&0.46 ~{\rm keV},  \nonumber \\
\Gamma _{f_{0}(1710)\rightarrow 2\gamma }(\Gamma _{f_{0}(1710) \rightarrow K%
\bar{K}}/\Gamma _{f_{0}(1710)tot}) &<&0.11~{\rm keV.}
\end{eqnarray}
Using the known branching ratio $\Gamma _{f_{0}(1500)\rightarrow 2\pi
}/\Gamma _{f_{0}(1500)tot}$ one gets $\Gamma _{f_{0}(1500)\rightarrow
2\gamma }<1.4$ ${\rm keV}${\rm \ }~\cite{Amsler:2002ey}. An accepted fit for 
$\Gamma _{f_{0}(1710)\rightarrow K\bar{K}}/\Gamma_{f_{0}(1710)tot}$ is not
reported in \cite{Eidelman:2004wy}. Using the value from~\cite%
{Longacre:1986fh} with $\Gamma _{f_{0}(1710)\rightarrow K\bar{K}}/ \Gamma
_{f_{0}(1710)tot}=0.38_{-0.13}^{+0.03}$ we find an upper limit of the order
of $\Gamma _{f_{0}(1710)\rightarrow 2\gamma }\sim 0.3$ keV.

For an absolute prediction of the two-photon decay widths we use $%
c_{e}^{s}=0.0761$ {\rm GeV}$^{-1}$ as determined in the model approach of
Ref.~\cite{Giacosa:2004ug}. For the non-strange quarkonium state we get $%
\Gamma _{N\rightarrow 2\gamma }=0.904$ keV, while for the isovector and
mixed scalars we have: 
\begin{eqnarray}
\Gamma _{f_{1}\rightarrow 2\gamma } &=&0.702~{\rm keV}\,,\Gamma
_{f_{2}\rightarrow 2\gamma }=0.258~{\rm keV}\,,  \nonumber \\
\Gamma _{f_{3}\rightarrow 2\gamma } &=&0.004~{\rm keV}\,,\Gamma
_{a_{0}^{0}\rightarrow 2\gamma }=0.345~{\rm keV}\,.
\end{eqnarray}%
The results for the mixed states are below the current upper limits.

The estimate for the $2\gamma $ decay of the bare quarkonium state $\bar{n}n$
of $0.904$ keV is smaller than the one of~\cite{Amsler:2002ey}, where the
following expression has been used: 
\begin{equation}
\Gamma _{N\rightarrow 2\gamma }(0^{++})=k\left( \frac{M_{N}(0^{++})} {%
M_{N}(2^{++})}\right) ^{3}\Gamma _{N\rightarrow 2\gamma }(2^{++})
\end{equation}
The coefficient $k$ is $15/4$ in a non-relativistic calculation, but becomes
smaller when considering relativistic corrections~\cite{Li:1990sx}. In~\cite%
{Amsler:2002ey} a range of values for $k$ from $2$ to $15/4$ is considered.
Our chiral Lagrangian approach combined with~\cite{Giacosa:2004ug} points to
a smaller value of $k$; using our result for $\Gamma _{N\rightarrow 2\gamma
}(0^{++})$ and taking the value $\Gamma _{N\rightarrow 2\gamma
}(2^{++})=2.60\pm 0.24$ keV~\cite{Eidelman:2004wy} at $M_{N}(2^{++})=1.27$
GeV we get $k\sim 0.25.$ This result is model dependent, since it relies on
the parameters for the covariant description of the scalar mesons used in~%
\cite{Giacosa:2004ug}. Hence a fully covariant treatment implies strong
deviations from the non-relativistic limit.

{\it Results when including glueball decays} - In line with strong coupling
arguments~\cite{Amsler:1995tu} direct glueball decay can be suppressed,
leading to the phenomenology discussed previously. Including explicit
glueball decay in a full fit to the data points of Table 1 with nine
parameters (three bare masses, two mixing parameters $f$ and $\epsilon$,
four decay strengths $c_{d}^{s},c_{m}^{s},c_{d}^{g},c_{m}^{g}$) does not
lead to a meaningful and unique $\chi^2$ minimum. To study the effect of
direct glueball decay in the context of an effective Lagrangian, we pursue
following strategy: we keep the bare masses listed in (\ref{fitpar}), which
are in accord with other phenomenological considerations, and only fit the
mixing parameters and decay strengths.

We first consider the $SU(3)$ flavor symmetry limit for the direct glueball
decay, that is $c_{m}^{g}=0$. A fit to the data of Table 1 results in the
parameter values: $f=0.0650$ GeV$^{2}$, $\varepsilon =0.211$ GeV$^{2}$, $%
c_{d}^{s}=8.50$ MeV, $c_{m}^{s}=2.58$ MeV and $c_{d}^{g}=-0.004$ MeV with $%
\chi ^{2}=29.00.$ The flavor symmetric glueball decay constant $c_{d}^{g}$
is very small in this fit, whereas the other parameters are practically
unchanged resulting in a $\chi ^{2}$ only sligthly smaller than before.

Flavor symmetry breaking in the direct glueball decay is encoded in the
strength parameter $c_{m}^{g}$. Inclusion of this additional parameter in a
full fit results in the set of parameters: $f=0.0654$ GeV$^{2}$, $%
\varepsilon =0.203$ GeV$^{2}$, $c_{d}^{s}=8.90$ MeV, $c_{m}^{s}=0.87$ MeV, $%
c_{d}^{g}=0.42$ MeV and $c_{m}^{g}=-1.17$ MeV with $\chi ^{2}=28.83$. The
dominant parameters $f,\varepsilon $ and $c_{d}^{s}$ remain stable, when
including the direct glueball decay. The coupling $c_{m}^{s}$ decreases
slightly, whereas the glueball coupling $c_{d}^{g}$ remains suppressed and
is smaller than $c_{m}^{g}$. Latter effect might suggest a strong flavor
symmetry breaking when discussing direct glueball decays. Again, the
resulting masses of the scalar-isoscalar mesons and decay rates listed in
Table 1 remain essentially unchanged for a pure fit to the data. Without
further guidance from theory concerning the direct glueball decay the
present phenomenological fit seems to indicate that this decay modes are
suppressed in line with the original arguments given in~\cite{Amsler:1995tu}%
. This result, however, contradicts the approach chosen in~\cite%
{Close:2001ga}, where direct glueball decays dominate.

In the context of an effective chiral Lagrangian, evaluated at tree level,
we discussed the phenomenological consequences for the scalar meson sector.
Thereby, the scalar glueball is assumed to mix minimally with the
scalar-isoscalar quarkonia states. A fit of the mixing parameters and
coupling streng\-ths entering in this scheme to the accepted data yields
several results: the direct flavor mixing between quarkonia states~\cite%
{Minkowski:2002nf} leads to a structure of the physical $f_0$ states, which
differs from the conventional scheme as in Refs.~\cite%
{Close:2001ga,Strohmeier-Presicek:1999yv}. The different pictures can be
directly tested in radiative $2\gamma$ decays of the $f_0$ mesons. The
relevance of direct glueball decays, which play a dominant role in~\cite%
{Close:2001ga}, cannot be verified in the present scheme. Further input from
lattice simulations, as for example demonstrated in Ref.~\cite{Sexton:1996ed}%
, could serve to obtain a better understanding of this issue.

{\bf Acknowledgments}

\noindent This work was supported by the DFG under con\-tra\-cts FA67/25-3,
GRK683. This research is also part of the EU Integrated Infrastructure
Initiative Had\-ron\-phy\-si\-cs project under contract number
RII3-CT-2004-506078 and President grant of Russia "Scientific Schools"
No.1743.2003.

\end{document}